\documentclass[prb,preprint]{revtex4-1} 

\usepackage{amsmath}  
\usepackage{amsfonts} 
\usepackage{graphicx} 

\newcommand{\beq}{\begin{equation}}
\newcommand{\eeq}{\end{equation}}
\newcommand{\nbd}{\nobreakdash}
\begin{document}


\title{Origin of the Zeroth Law of Thermodynamics and its Role in Statistical Mechanics}

\author{Kim Sharp}

\date{\today}

\begin{abstract}

In statistical mechanics the zeroth law of thermodynamics is taken as a postulate which, as its name indicates, logically precedes the first and second laws. Treating it as a postulate has consequences for how temperature is introduced into statistical mechanics and for the molecular interpretation of temperature. One can, however, derive the zeroth law from first principles starting from a classical Hamiltonian using basic mechanics and a geometric representation of the phase space of kinetic energy configurations - the velocity hypersphere. In this approach there is no difficulty in providing a molecular interpretation of temperature, nor in \textit{deriving} equality of temperature as the condition of thermal equilibrium. The approach to the macroscopic limit as a function of the number of atoms is easily determined. One also obtains with little difficulty the Boltzmann probability distribution, the statistical mechanical definition of entropy and the configuration partition function. These relations, along with the zeroth law, emerge as straightforward consequences of atoms in random motion.
\end{abstract}

\maketitle 
%
\section{Introduction} 
The first law of thermodynamics, conservation of energy, was discovered through the work of Rumsford  \cite{rumsford_1798}, Mayer \cite{mayer_1842}, Joule \cite{joule_1850} and others. Discovery of the first law was contemporaneous with the development of thermodynamics \cite{dugdale_entropy_1966}. The second law of thermodynamics has its origin in the work of Carnot \cite{carnot1872reflexions}. Clausius coined the term entropy and gave it a precise thermodynamic definition \cite{clausius_modified_1854, clausius_several_1865}. This led to various formulations of the second law including ``Heat can never pass from a colder to a warmer body without some other change, connected therewith, occurring at the same time.'' \cite{clausius_modified_1854}, or equivalently, ``It is impossible for a self-acting machine, unaided by any external agency, to convey heat from one body to another at a higher temperature." \cite{kelvin_1851}. The second law was discovered within the framework of classical thermodynamics, which is agnostic about the nature of matter, energy, heat and temperature: Matter and energy are treated as continua, and temperature is operationally defined as that which is measured by an ideal (gas) thermometer. But matter actually consists of atoms and heat is the random motion of these atoms. The great contributions of Boltzmann \cite{boltzmann_uber_1877,sharp_translation_2015} and Gibbs \cite{gibbs_elementary_1902} were to provide definitions of entropy and the second law in terms of the statistical distributions of these atoms' positions and velocities. After the development of quantum theory statistical mechanics (more accurately statistical thermodynamics) was reformulated on the basis of the discrete energy levels (quantum states) available to the system. While this is in principle a more fundamental description than classical statistical mechanics it is somewhat ironic that one still has to appeal to classical thermodynamics, a theory that has no atoms in it, to put the `thermo' into statistical thermodynamics. Specifically, to quote Fowler and Guggenheim \cite{fowler_guggenheim_1939} \textit{``If two assemblies are each in thermal equilibrium with a third assembly, they are in thermal equilibrium with each other. From this it may be shown to follow that the condition for thermal equilibrium between several assemblies is the equality of a certain single-valued function of the thermodynamic states of the assemblies, which may be called the temperature T, any one of the assemblies being used as a thermometer... This postulate of the ``existence of temperature'' could with advantage be known as the zeroth law of thermodynamics.''} They named it the zeroth law to indicate its logical precedence over the other two laws.

The zeroth law postulate is, to the author's knowledge, relied upon in all the most widely used statistical mechanics textbooks. A well known introductory textbook states at the outset that ``Temperature is the thing that's the same for two objects, after they've been in contact long enough" \cite{schroeder_thermalphysics_2000}. A representative way the zeroth law is used is as follows \cite{hill_introduction_1962, reif_fundamentals_1965, mcquarrie_statistical_1976, callen_thermodynamics_1985, schroeder_thermalphysics_2000}: For two systems $A$ and $B$ in thermal contact (able to exchange heat), one maximizes the number of microstates of the combined system subject to constraints on the numbers of atoms, $N_A$ and $N_B$, and the total energy $E = E_A + E_B$.  (In a fully quantum treatment quantum states and microstates are synonymous.)  This condition is satisfied mathematically when a certain parameter, often given the symbol $\beta$, assumes the same value in each system, namely $\beta_A = \beta_B$. Typically the two $\beta$'s are introduced as Lagrangian multipliers to facilitate the maximization. Now, assuming the zeroth law one deduces that $\beta$ must be some monotonic but yet undetermined function of temperature so that $\beta(T_A) = \beta(T_B)$ implies $T_A = T_B$. The classical thermodynamics relation $1/T = \partial S/ \partial U$ can be used to make the assignment $\beta=1/k_bT$, where $S$ is the entropy, $U$ is the internal energy, and $k_b$ is the Boltzmann constant. One can derive expressions for the partition function and the three important statistical mechanics distributions, the Boltzmann, Fermi and Bose-Einstein distributions, using $\beta=1/k_bT$ without ascribing any further meaning to $T$. Another fundamental concept that is dependent on the zeroth law postulate is that of a thermal reservoir, which keeps the system of interest at a constant temperature. The thermal reservoir concept is used to define the canonical ensemble, \cite{mcquarrie_statistical_1976, callen_thermodynamics_1985} and it is an alternative way to interpret the parameter $\beta$ \cite{tolman_principles_1938}.

This reliance on the zeroth law assumption and the \textit{thermodynamic}  definition of temperature is not without consequences, however. First, there is no immediate interpretation of $T$ in molecular terms. Indeed, in one textbook it is only five chapters later that, by use of the classical approximation and the equipartition theorem, it is shown that temperature is proportional to mean kinetic energy per atom,  $T =  2\bar E/3k_b$ \cite{mcquarrie_statistical_1976}. Second, logical deduction from the relation  $1/T = \partial S/ \partial U$ leads one to conclude that certain systems under special conditions have a negative temperature and thereby a negative heat capacity, $C_v$.  \cite{braun_negative_2013} This interpretation is controversial \cite{dunkel_consistent_2014}, and it results from applying a definition of $T$ derived from the equilibrium, continuum model of classical thermodynamics to an atomistic non-equilibrium system having an inverted distribution among energy states, i.e. one where higher energy states are more populated than lower energy states. Morover, the inference from negative $T$ to negative $C_v$ is inconsistent with the alternate statistical mechanical definition of $C_v$ derived from the mean square fluctuation of a system's energy in the canonical ensemble  \cite{mcquarrie_statistical_1976}, since of course the square of the fluctuations can never be negative. Third, one can hardly say that statistical thermodynamics is derived from first principles, even if one starts from quantum mechanics, if one has to borrow that quintessential thermal quantity temperature from classical thermodynamics, a theory with no atoms in it. One might attribute this difficulty to the fact that there is no quantum operator for $T$. Finally, there are many areas of statistical mechanics where the treatment is classical from the outset: Theories of polyelectrolytes, polymers, macromolecules, simulations of complex multicomponent systems. These more often than not start with a classical Hamiltonian. In that case, as is shown here, one need not assume the zeroth law as a postulate. It can be \textit{derived} from first principles using just the minimal concepts of mechanics: velocity, kinetic energy, potential energy and a geometric construct, the velocity hypersphere, to represent the phase space of kinetic energy micro-states. Temperature is from the start simply proportional to the mean kinetic energy per atom. So $T$ can never be negative. While this derivation recapitulates the familiar statistical mechanical relations such as the Boltzmann factor and the configuration partition function, the derivation is likely new to those who, like the author, learned statistical mechanics from the standard textbooks. This approach also makes it easy to see how rapidly these relations approach the large-N or macroscopic limit. This is of interest since the number of atoms in current computer simulations is still many orders of magnitude fewer than Avogado’s number. 
\section{Theory}

%
\subsection{Macrostates and  Microstates}
In principle, if one knew the state of each atom in a system (the so called microstate) at all times one could obtain the macroscopic properties of the system by taking suitable averages over the microstates.  Given the enormous number of atoms contained in even a tiny amount of matter, explicit consideration of every microstate is impossible.  Instead, one describes the system in terms of statistical distributions. To this end one must first define a microstate and then be able to enumerate the number of microstates available to the system subject to any constraints on it. 

In a classical treatment the microstate of a system with $N$ atoms is represented by a point in the 6N-dimensional position-momentum phase space
\beq
\label{eq:microstate}
\textbf{X} = (x_1,y_1,z_1,a_1,b_1,c_1 \ldots x_N,y_N,z_N,a_N,b_N,c_N)
\eeq
where $N$ is the number of atoms, $x_i, y_i, y_i$ are the coordinates of the $ i^{th}$ atom and $a_i, b_i, c_i$ are the $x, y, z$ components of its velocity. The coordinates alone will be referred to as a spatial configuration/microstate. A set of velocities  $(a_1 \ldots c_N)$ will be referred to as a velocity microstate, or equivalently, a kinetic energy microstate. The combined set of coordinates and velocities will be referred to as a spatio-kinetic microstate.

The total energy of the system is given by the classical Hamiltonian, which is the sum of kinetic and  potential energies: $E = E_k + U$. The kinetic energy is a sum over $3N$ velocity terms
\beq
E_k = \sum_{i=1}^N  \,^1/_2 m_i(a_i^2 + b_i^2 + c_i^2)
\eeq
where $m_i$ is the mass of the $i^{th}$ atom. The potential energy is a function of the 3N spatial coordinates, 
\beq
\label{eq:Upot}
U =  U(x_1,y_1,z_1 \ldots x_n,y_n,z_n).
\eeq
The force on an atom is given by the negative gradient of $U$ with respect to that atom's position.

%
\subsection{Counting the number of kinetic energy microstates}
To determine the number of ways to partition a given amount of kinetic energy among a set of atoms consider a single atom of mass $m$ in the absence of any forces. If the components of this atom's velocity in the $x$, $y$, and $z$ directions at some instant are $a$, $b$ and $c$, respectively, then the magnitude of the velocity  (speed) is $v = (a^2+b^2+c^2)^{1/2}$ and the atom's kinetic energy is
\beq
E_k =  \,^1/_2  mv^2 = \,^1/_2 ma^2 + \,^1/_2 mb^2 +\,^1/_2 mc^2.
\eeq 
The velocity of this atom can be represented graphically by a vector, or a point $p$ with coordinates $(a,b,c)$ in three dimensional ($3D$) space, Fig. 1. 
\begin{figure}[h!]
\centering
\includegraphics[width=5in]{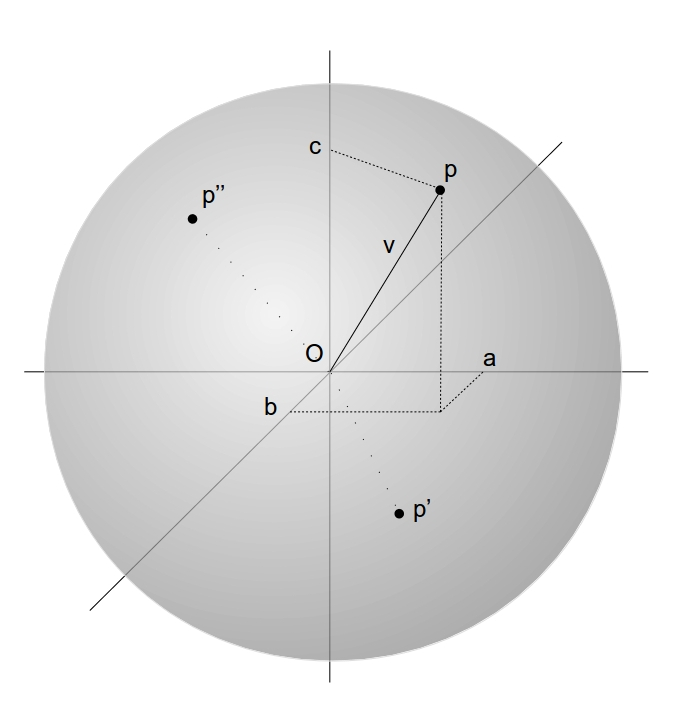}
\caption{Velocity sphere for a single atom with speed $v$ and  kinetic energy $E_k = \,^1/_2  mv^2$. Three possible combinations of velocity components are represented by the points $p$, $p'$ and $p''$ lying on the surface of a sphere of radius $R = (2E_k/m)^{1/2}$}
\label{fig:hsphere}
\end{figure}
Of course there are many different combinations of velocity components $(a',b',c')$ with the same total kinetic energy, providing these components satisfy  $v = (a'^2+b'^2+c'^2)^{1/2}$.  In fact every possible partition of velocity components with kinetic energy $E_k$ corresponds to a point on the surface of a sphere of radius $R = (2E_k/m)^{1/2}$. The surface of this sphere thus represents the surface of constant kinetic energy in the velocity coordinate space. The number of ways to partition this amount of kinetic energy, $W$, is proportional to the surface area of the sphere, so
\beq
\label{eq:ways_1atom}
W \propto R^2 \propto (E_k^{1/2})^2.
\eeq
Here the exponents $^1/_2$ and $2$ have not been combined for a reason which will become apparent upon generalization to $N$ atoms. Note that this proportionality is the same as that obtained for the density of states in the Sackur-Tetrode derivation of the entropy of an ideal gas, the counting arguments being very similar.  \cite{hill_introduction_1962, mcquarrie_statistical_1976} The difference is that here the phase space is the $x,y,z$ velocity components of a single atom, not the three quantum numbers of a particle in a box. 

Now consider $N$ atoms of mass $m$ inside a rigid container of volume $V$ with perfectly reflecting walls. The total kinetic energy is $E_k$, and there are no forces acting between or on the atoms. This corresponds to an ideal mono-atomic gas under conditions of constant $N, V$ and $E$, referred to in the standard approach as the microcanonical ensemble.  The atoms are in rapid random motion, colliding elastically with each other and the walls of the container, their positions and velocities incessantly changing with time according to the laws of motion. The generalization of Eq. (\ref{eq:ways_1atom}) proceeds using the same geometric analogy. $E_k$ is constant, so at every instant the $3N$ velocity components $(a_1,b_1,c_1, \ldots a_N,b_N,c_N)$ must satisfy $E_k = \, ^1/_2  m \sum^N_i (a_i^2 + b_i^2 + c_i^2)$.  Each possible combination of velocity components can be represented as a point with coordinates $(a_1,b_1,c_1, \ldots a_N,b_N,c_N)$  in 3N\nbd-dimensional space, lying a constant distance  
\beq
\label{eq:nvector}
R = \left( \sum^N_{1} (a_i^2 + b_i^2 + c_i^2)\right)^{1/2} = \, (2E_k/m)^{1/2}
\eeq
from the origin. These points define the surface of a $3N$\nbd-dimensional hypersphere of radius $R = (2E_k/m)^{1/2}$, which is referred to as the velocity hypersphere. The surface of this hypersphere has dimension $3N-1$, one less than the velocity coordinate degrees of freedom because of the constant energy constraint. The number of different ways to partition the kinetic energy $E_k$ among the atoms is proportional to the area of the constant energy surface, i.e. the area of a $3N$\nbd-dimensional hypersphere, which scales as $R^{(3N-1)}$.  \cite{nist_digitallibmath} Following Eq. \ref{eq:ways_1atom} this gives the scaling law
\beq
\label{eq:ways_Natom}
W  \propto E_k^{(3N-1)/2}.
\eeq
As will become apparent, only changes in $W$ with respect to the kinetic energy are needed to derive the basic statistical mechanical relations. Any multiplicative constants, including the atomic mass, will cancel. A mixture of atoms of different masses results in a hyper-ellipsoidal surface of constant energy. It is shown in Appendix \ref{sub:masseffect} that this does not change the form of the scaling law, Eq. (\ref{eq:ways_Natom}).

The next step is to determine how the number of ways to partition the kinetic energy among the atoms changes when $E_k$ is increased by a small amount $\delta E_k$. The total kinetic energy is now $E_k + \delta E_k$, with a corresponding increase in the radius of the velocity hypersphere to $R' = (2(E_k + \delta E_k)/m)^{1/2}$. From Eq. (\ref{eq:ways_Natom}) the ratio of the number of ways to partition the kinetic energy before and after the increase is 
\beq
\label{eq:wayratio1}
\frac{W'}{W} = \left({1+  \frac{\delta E_k}{E_k}} \right)^{(3N-1)/2}.
\eeq
Writing this equation in logarithmic form, and setting $3N-1 \rightarrow 3N$ since $N$ is very large, we obtain
\beq
\label{eq:lnWays}
\ln ({W'}/{W} )= \frac{3N}{2} \ln \left(1 + \frac{ \delta E_k}{E_k} \right).
\eeq
If the change in kinetic energy is small compared to the total we can use the approximation $\ln(1+x) \approx x$ when $x<<1$  to write Eq. (\ref{eq:lnWays}) as $\ln(W'/W) = \frac{3N}{2}\frac{\delta E_k}{E_k}  $, or in exponential form
\beq
\label{eq:boltz1}
\frac{W'}{W} = \exp \left( \frac{3 \delta E_K}{2\bar{E}_k} \right),
\eeq
where $\bar{E}_k = E_k/N$ is the mean kinetic energy per atom.

The change in kinetic energy in Eq. (\ref{eq:boltz1})  could be produced by adding a small amount of heat $\delta q$ to the gas, by doing some mechanical work $\delta w$ on the gas by compressing it, or by some combination of the two. At this point, therefore, it is useful to clarify the concept of heat as used here, and avoid upsetting thermodynamic purists.  Heat is defined in purely mechanical terms: it is simply the random motion of atoms. The change in the amount of heat in a non-rotating  body at rest is just the change in its total kinetic energy $ \delta E_k$, however this is produced.  It is also meaningful to say that the total amount of heat in that body is its total kinetic energy, $E_k$, although nothing here relies on the ability to quantify the total amount of heat. 

Equation (\ref{eq:boltz1}) is the fundamental relation from which the other key relations will be derived. The extremely high dimensionality of the hypersphere \--- $N$ is  of order $10^{18}$ or more for macroscopic systems \--- imparts an exponential dependence of area on radius, leading to the exponential dependence of $W$ on $\delta E_k$. The increase in $W$ is independent of the number of atoms and their masses and it depends only on the ratio of the increase in kinetic energy to the mean kinetic energy per atom. Thus, a given increment in $E_k$ increases $W$ by a larger factor when the mean kinetic energy per atom is lower. This behavior is crucial for understanding thermal equilibrium.
%
\subsection{Accounting for forces between atoms}
\label{sec:forces}
\begin{figure}[h!]
\centering
\includegraphics[width=5in]{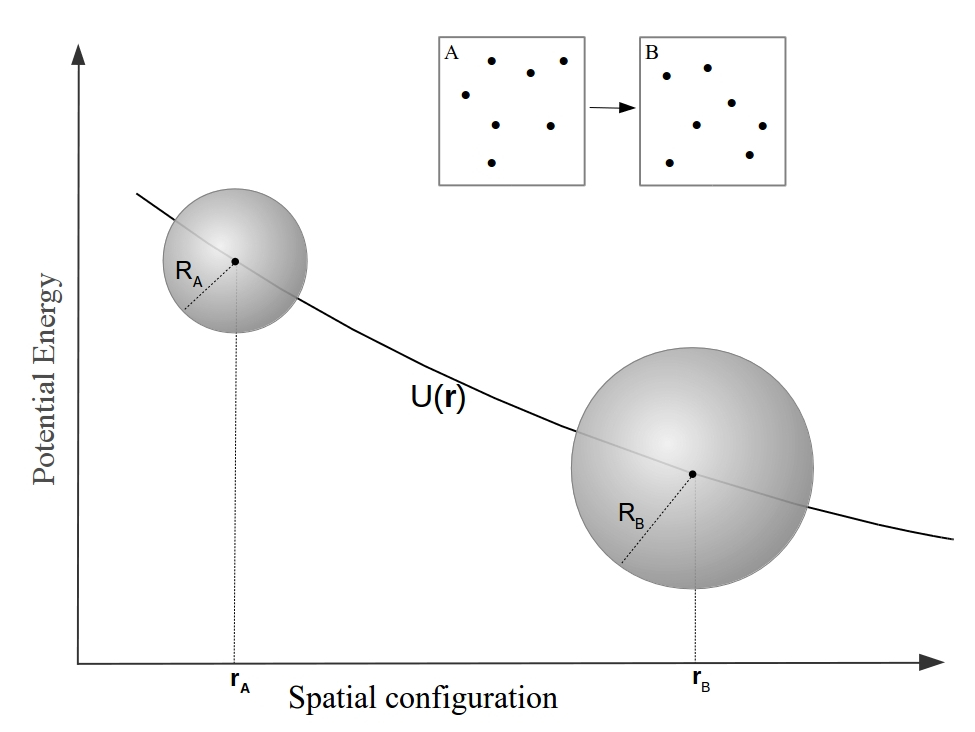}
\caption{Two spatial configuration of atoms, $A$ and $B$, are illustrated schematically in the upper boxes. Forces between atoms produce a $3N$\--dimensional potential energy surface over which the atoms move, $U(\textbf{r})$. Configurations  $A$ and $B$ differ in total potential energy, indicated on the vertical axis. $3D$ slices through the $3N$\--dimensional velocity hyper-spheres corresponding to configurations  $A$ and $B$ are depicted by the spheres. Since $B$ is the spatial configuration with lower potential energy it has more kinetic energy, thus a larger velocity hypersphere and consequently more ways to partition the kinetic energy among the $N$ atoms than configuration $A$.}
\label{fig:sphereforce}
\end{figure}

Equation (\ref{eq:boltz1})  is now applied to a system where there are forces acting on the atoms. Consider $N$ atoms in a perfectly insulating and rigid container (constant $NVE$  conditions.)  Two different instantaneous spatial configurations of these atoms, $A$ and $B$, are indicated schematically in Fig. \ref{fig:sphereforce}.
If there are forces acting on the atoms then in general the potential energies of the two configurations, $U_A$ and $U_B$, will be different. Since total energy is conserved upon moving from configuration $A$ to configuration $B$, this change in potential energy is converted to kinetic energy if $U_B < U_A$  (heat is generated), or kinetic energy is converted to potential energy if $U_B > U_A$ (heat is absorbed). The mechanism of energy conversion is simply Newtonian dynamics: The acceleration/deceleration of each atom as it moves down/up a potential energy gradient.  If the difference in potential energy is $\Delta U = U_B -  U_A$, then $\delta E_k = -\Delta U$. Applying Eq. (\ref{eq:boltz1}) the ratio of the number of ways to partition the kinetic energy among the atoms for spatial configurations $B$ versus $A$ is
\beq
\label{eq:wayratio2}
\frac{W_B}{W_A} = \exp \left( - \frac{3\Delta U}{2\bar{E}_k}\right).
\eeq
If $\Delta U$ is negative there are more ways to partition the kinetic energy among the atoms in configuration $B$ because there is more of it. The converse is true if $\Delta U$ is positive. Using the postulate that all permissible microstates are equally likely, \cite{tolman_principles_1938}  the ratio of probabilities of any two spatial configurations $A$ and $B$ is 
\beq
\label{eq:boltz3}
\frac{p_B}{p_A} = \exp  \left( - \frac{3\Delta U}{2\bar{E}_k}\right), 
\eeq
which produces the Boltzmann distribution. 
%
\subsection{The Boltzmann distribution and observable states}

Since Eq. (\ref{eq:boltz3})  is derived from Eq. (\ref{eq:boltz1}) it is accurate under the same conditions, namely when $\Delta U << E_k$.  Put another way, the Boltzmann distribution is followed when the difference in mean kinetic energy between two spatial configurations is negligible. As will be shown, when $N$ is large this is true for any pair of \textit{observable states}.   Let the total kinetic energy in the spatial configuration with the lowest potential energy $o$ be $E^o_k$. Let the potential energy of spatial configuration $i$ be higher by some multiple of the mean kinetic energy of configuration $o$, $\Delta U_i = c \bar{E}_k^o$, where $c$ is positive. Then the total kinetic energy in configuration $i$ is lower by the same amount,  so its mean kinetic energy is lower: 
\beq
\bar{E}_k^i = \bar{E}_k^o - c \bar{E}_k^o/N,
\eeq
and the relative probability of observing configuration $i$ relative to configuration $o$ is
\beq
\frac{p_i}{p_o} \approx \exp \left( - \frac{3c}{2(1-c/N)}\right)
\eeq
If the factor $c$ is greater than a few orders of magnitude, then the probability of observing $i$ is negligible. Conversely, if $c$ is small enough for $i$ to be observable with significant probablity, then the mean kinetic energy differs negligibly from $\bar{E}_k^o$ because $N$ is very large .

A numerical example demonstrates the force of this argument. Let $N = 10^{12}$, which is still a very small amount of matter. Say the potential energy of a spatial configuration is high enough relative to configuration $o$ to lower the mean kinetic energy by just one part in a hundred million so $c/N = 10^{-8}$. Therefore $ \Delta U_i/\bar{E}^o_k =10^{4}$ and the probability of observing this configuration is of order $e^{-10000}$ times lower than that of the lowest energy configuration, which is negligible. From this we extract two equilibrium properties of a system of interacting atoms under conditions of constant N,V,E:

Property I: The set of spatial configurations that are observable - that contribute to macroscopic properties - differ negligibly in their mean kinetic energy. Within this extremely narrow band of mean kinetic energies, the probability distribution of configurations will follow almost exactly the Boltzmann distribution. 

Property II:  This set of observable configurations will make the overwhelming contribution to the total number of kinetic energy micro-states of the system. 
%
\subsection {Thermal Equilibrium}
\label{sec:equilibrium}
The conditions for equilibrium of any two bodies in thermal contact is now derived. For this Eq. \ref{eq:boltz1} must be generalized to include forces acting between the atoms. Consider again the system of $N$ atoms under conditions of constant $V$, discussed in section \ref{sec:forces}. A small amount of heat $\delta E_k << {E_k}$ is added.  Since all the observable spatial configuration have the same value of $\bar{E_k}$ (Property I above) the number of ways to distribute the kinetic energy increases by the same ratio in each of these configurations. Thus
\beq
\label{eq:bfactor1}
\frac{W'_i}{W_i} =  \exp{\left(\frac{3\delta E_k}{2\bar{E}_k}\right)}.
\eeq
Since these observable configurations contribute the overwhelming number of kinetic energy microstates (Property II above) the ratio of the total number of spatio-kinetic energy microstates before and after heat addition increases by the same factor. Thus
\begin{equation}
\label{eq:waystotal1}
\frac{W'_{tot}}{W_{tot}} \approx 
\frac{\sum^{O}_i  W_i'}{\sum^{O}_i  W_i}  =
\exp{\left(\frac{3\delta E_k}{2\bar{E}_k} \right)} ,
\end{equation}
where the sums can be restricted to the $O$ observable spatial configurations. This factor is identical to that derived in the absence of forces (Eq. \ref{eq:boltz1}), demonstrating its general validity. 

Now take two bodies $A$ and $B$, each surrounded by a perfectly insulating rigid container, so no heat can flow in or out from the surroundings and no mechanical work can be done. Let their mean kinetic energies per atom be different, with $\bar{E_k}_{,B} > \bar{E_k}_{,A}$. Bring $A$ and $B$  into thermal contact, separated by a thin rigid diathermal wall. Let a small amount of heat $\delta q$ be transferred from $B$ to $A$. The total number of ways to partition the kinetic energy among all the atoms is the product of the number of ways to partition it in A and in B. Using Eq. \ref{eq:waystotal1} the change in total number of ways is
\beq
\label{eq:tequil}
 \frac{W'_{tot}}{W_{tot}} =\frac{ (W'_{tot,A} W'_{tot,B})}{(W_{tot,A} W_{tot,B})} = 
\exp \left({\delta q} \frac{3}{2}\left( \frac{1}{\bar{E_k}_{,A}} - \frac{1}{\bar{E_k}_{,B}} \right) \right)
\eeq
where  the heat changes for A and B are of equal magnitude but have opposite signs.  Since body $B$ has a greater mean kinetic energy the second factor in the exponent of Eq. (\ref{eq:tequil}) is positive. If heat $\delta q$ is also positive the exponent is positive and the total number of ways to partition the kinetic energy increases.  So this is more likely to be observed. Conversely, if heat flows in the opposite direction $\delta q < 0$ and the exponent is negative, corresponding to a decrease the total number of ways. This is less likely to be observed. Only when $\bar{E_k}_{,B} = \bar{E_k}_{,A}$ is no further increase in $W$ through heat flow possible. Equation (\ref{eq:tequil}) makes it clear that thermal equilibrium is established by reaching the macrostate of maximum likelihood, the one that can be realized by the most microstates. This is achieved when the mean kinetic energies are equalized. Note that in obtaining these two equivalent conditions, it was not assumed \textit{a priori} that temperature was equalized, or indeed that any quantity was equalized in the two bodies.
\subsection {The statistical mechanical temperature}
\label{sec:temperature}
It is well known that the pressure of an ideal monoatomic gas is given both by the ideal gas law $P = \rho k_b T$ and by the mechanical equation $P =\frac{ 2 \rho \bar{E}_k}{3}$, where $T$ is temperature, $k_b$ is Boltzmann's constant and $\rho$ is the gas density. If an ideal gas thermometer were placed in contact with either of the above two bodies $A$ and $B$, which are in thermal equilibrium, by application of Eq. (\ref{eq:tequil}) to each the ideal gas would attain the same mean kinetic energy per atom and therefore measure the same temperature $T = P/\rho k_b = \frac{2\bar{E}_k}{3k_b}$. The second condition for thermal equilibrium can now be restated as equalization of temperature, defined \textit{statistical mechanically} as the mean kinetic energy per atom (times a constant which is essentially just a units conversion factor.)  Replacing $2\bar{E}_k/3$ by $k_b T$ in Eq. (\ref{eq:tequil}) the zeroth law equation for thermal equilibrium is
\beq
\label{eq:tequilT}
 \frac{W'_{tot}}{W_{tot}} =
\exp \left(\frac{\delta q}{k_b} \left( \frac{1}{ T_A} - \frac{1}{T_B} \right) \right)
\eeq
Equation (\ref{eq:waystotal1}), the general equation for the increase in total number of spatio-kinetic microstates due to an increase in thermal energy,  becomes 
\begin{equation}
\label{eq:waystotal1T}
\frac{W'_{tot}}{W_{tot}} = \exp{\left(\frac{\delta E_k}{k_b T} \right)} , 
\end{equation}
%
\subsection{ The total number of microstates}
\label{sec:gamma_total}
A microstate is specified by both the spatial configuration and the partition of kinetic energies, $6N$ coordinates in all (Equation (\ref{eq:microstate})). Under $NVE$ conditions the number of ways to distribute the kinetic energy when the atoms are in spatial configuration $i$ is, using  Eq. \ref{eq:wayratio2} with $2\bar{E}_k/3$ replaced by $k_b T$
\beq
\label{eq:waysref}
W_i= W_o \exp \left(-\frac{U_i}{k_b T} \right).
\eeq
Here $W_0$ is the number of ways to distribute the kinetic energy in some reference state spatial configuration, chosen for mathematical convenience where $U=0$.  The reference state assumption is discussed in the Appendix. 

If the volume is discretized into small cubic cells of volume $b$, the  number of positions available to one atom $V/b$. Specifying the positions of all N atoms to the same resolution, the total number of spatial configurations is $M = (V/b)^N$.  The total number of spatio-kinetic microstates is the sum over all spatial configurations of the number of kinetic energy distributions for each spatial configuration:
\beq
\label{eq:waystotal}
W_{tot} = \sum^M_i W_i = W_0 \sum^M_i {\exp  \left(-\frac{U_i}{k_b T}\right)}.
\eeq
The value of $W_0$ is not known,  nor therefore is the absolute value of $W_{tot}$,  but this is not necessary as only ratios of total numbers will be needed. The total number of microstates will also depend on how finely space is discretized, but again, providing this is done finely enough this will not affect ratios of $W_{tot}$.

The normalized Boltzmann probability distribution can be written, using Eq.  \ref{eq:boltz3} with $2\bar{E}_k/3$ replaced by $k_b T$, as
\beq
\label{eq:boltz4T}
p_i = \frac{{\exp  \left( - {U_i}/{k_b T}\right)}}{Z},
\eeq
where the normalization factor
\beq
\label{eq:partition1}
Z = {\sum^M_i {\exp  \left(-\frac{  {U_i}}{k_b T} \right)}}
\eeq
is known as the configuration partition function.  We see from Eq. (\ref{eq:waystotal}) that the total number of spatio-kinetic configurations is proportional to the configuration partition function,
\beq
\label{eq:waystotalT}
W_{tot} = W_0 Z,
\eeq
Moreover, $Z$ describes how $W_{tot}$ depends on  the available spatial configurations, their potential energy and the temperature, hence the central importance of $Z$ in statistical mechanics.
%
\subsection{Calculating Macroscopic quantities}
\label{sec:averages}
One of the goals of statistical mechanics is to obtain macroscopic averages, especially observables, from microscopic behavior. The macroscopic average value of any quantity $Y$ can now be obtained using the normalized probability given by Eq. (\ref{eq:boltz4T}) as
\beq
\label{eq:averages}
< Y > = \sum_i^M p_i Y_i,
\eeq
provided its value $Y_i$ can be defined for any spatial microstate $i$.  Two of the most important quantities are  the average potential energy and its mean squared fluctuation, given by  ${<}U{>} = \sum_i^M p_ i U_i$ and   ${<}\delta U^2{>} = \sum_i^M p_i (U_i  -{ <}U{>})^2$, respectively.   

It is necessary that the discretization is fine enough that averages given by Eq. (\ref{eq:averages}) are independent of the value of $b$.  Provided $b$ is small enough that the variation in $U$ across a cell is small compared to $\bar{E}_k$ this will be the case. If the discretization volume $b$ is reduced further,  there will be more spatial configurations in the sum, but also more configurations in the sum for $Z$, so $p_i$ values will decrease proportionally and the macroscopic average will be unchanged.

\subsection{Definition of Entropy}
As a formal  treatment of equilibrium statistical mechanics it would be possible to stop here at the point of counting microstates. The Boltzmann distribution and configurational partition function have been obtained. The conditions for thermal equilibrium have been established. Any equilibrium macroscopic quantity can in principle be calculated from Eq. (\ref{eq:averages}). However, prior to the full development of statistical mechanics a thermodynamic quantity entropy ($S$) had been defined, the increase in $S$ was the essence of the second law of thermodynamics, and  maximization of $S$ was shown to be the condition for thermal equilibrium.  \cite{clausius_several_1865} Clausius showed that when a small amount of heat $\delta q$ is added to a body  under `reversible' conditions the entropy change is given by $\delta S = \delta q/T$. \cite{clausius_modified_1854} This is just the term in the exponent of Eq. (\ref{eq:waystotal1T})  for the change in the total number of ways to distribute the heat. A more accurate term than 'reversible' is quasistatic, meaning i) The amount of heat is small compared to the total kinetic energy, so $T\approx$ constant. ii) There is time for the added heat to be distributed through the body. In other words, it becomes accurate to say that on average the kinetic energy of each atom is raised by the same amount. Under these conditions  Eq. (\ref{eq:waystotal1T}) may be re-written as
\beq
\label{eq:slnw1}
\delta S = k_b \ln(W'_{tot}/W_{tot})
\eeq
for a change in entropy. Note that the expression  $\delta S = \delta q/T$ was derived from thermodynamics, meaning that it is valid for the addition of heat to any type of substance: Whatever kind of atoms it is composed of, or whatever kind of forces are acting between the atoms. Similarly, Eq. \ref{eq:waystotal1T}  has the same general validity, as shown in section \ref{sec:forces}.  For the absolute entropy 
\beq
\label{eq:slnw2}
 S = k_b \ln(W_{tot}) + constant,
\eeq
which is Boltzmann's famous equation \cite{planck_section_1906}.  Boltzmann was the first to discover this link between the thermodynamic and statistical definitions of entropy through $W_{tot}$, the number of microscopic states by which a given macro-state can be realized. \cite{boltzmann_uber_1877} In almost all situations of practical interest we are concerned only with changes in entropy and so we do not need the value of the constant in Eq. (\ref{eq:slnw2}). Referring back to Eq. (\ref{eq:tequilT}) we can now add a third equivalent condition for thermal equilibrium: It is the state of maximum entropy. 

Following  Lebowitz's lucid summary  \cite{lebowitz_boltzmann_1993}, we can now explain the second law of thermodynamics as follows. The system has a given number of spatio-kinetic microstates available to it, $\Gamma$, determined by the composition of the system and any external constraints on it. The fundamental postulate of statistical mechanics is that the system is equally likely to be found in any one of these microstates. \cite{tolman_principles_1938}  A macrostate is defined by a fairly small number of measurable bulk or macroscopic quantities such as volume, density, pressure, temperature.  All of the microstates that produce a given set of macroscopic or bulk quantities (within experimental error) can be said to belong to the corresponding macrostate $J$.  From the fundamental postulate it follows that this macrostate will be observed with a probability proportional to its number of microstates $\Gamma_J$. A system that is not at equilibrium will likely evolve by moving through a succession of macrostates with increasing  $\Gamma_J$ until it reaches the macrostate $J^*$ having the maximum number of microstates. \cite{boltzmann_uber_1877} This will be the equilibrium state, the one that can be realized by the most microstates. Using Eq. (\ref{eq:slnw1}) the difference in entropy between any two macrostate $J$ and $J'$  is $\Delta S = \ln  \Gamma_{J'}/\Gamma_J$, so we can rephrase the evolution of a non-equilibrium system thus: It will likely evolve by moving through a succession of macrostates with increasing  entropy until it reaches the state of maximum entropy.  Furthermore, as is apparent if one estimates numbers of microstates for a typical macroscopic system (See below), $\Gamma_J$ increases at such a stupendous rate as one moves towards the equilibrium macrostate that i) the direction of evolution is effectively deterministic, not probabilistic; ii) after sufficient time for the equilibrium state to be reached, macrostate $J^*$ or ones that are indistinguishably close to it are the only ones that will be observed from then on. In this manner irreversibility arises at the macroscopic level. 
%
 \subsection{Approach to macroscopic behavior}
\begin{figure}[h!]
\centering
\includegraphics[width=5in]{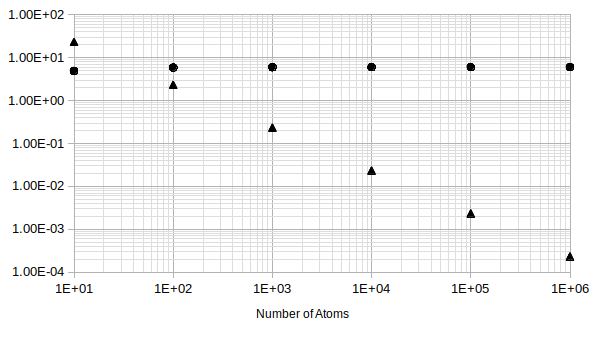}
\caption{Entropy change in units of $k_b$ upon adding $4 \bar E_k$ of heat to a system with an increasing number of atoms (circles). Limiting value at large N is $6 k_b$. \% difference in entropy change relative to the large N limit (triangles)}
\label{fig:dS_vs_N}
\end{figure}

%
Expressions for the Boltzmann distribution, the configuration partition function, entropy and the 'zeroth law' equation were derived assuming that the number of atoms was very large. How rapidly is this 'macroscopic limit' approached as a function of N? This is easily determined using  Eq. (\ref{eq:wayratio1}), which does not assume that the number of atoms is large or that the amount of heat added is small compared to the total kinetic energy. This equation was used to calculate the increase in number of ways to partition kinetic energy when $4 \bar E_k$ of heat is added to a system containing between $10$ and $1000,000$ atoms, and from this the change in entropy was calculated. The difference from the limiting value at large N -- the macroscopic limit -- was also calculated using Eq. \ref{eq:waystotal1}. The results are plotted in Fig. \ref{fig:dS_vs_N}.  Already at a thousand atoms the error in the entropy change is only about $0.2\%$. The error then decreases by a factor of $10$ for every factor of $10$ increase in $N$.  This implies that the error in bulk  themodynamic properties from early molecular simulations, which could handle only a few hundred particles  \cite{pollack_alder_1978}, was not large. Such simulations are now routinely done with $10^5 - 10^6$ atoms, where the results are expected to differ negligibly from the macroscopic limit.
%

\subsection{Accuracy of the Zeroth Law}
%
Statistical thermodynamics is by its nature atomistic and statistical, and so it allows for the possibility of fluctuations away from equilibrium. It is therefore of interest to see how accurately the zeroth law condition is obeyed in a representative case.  Take two $1$g masses of water, $A$ and $B$,  in thermal contact. Initially they are at equilibrium with $T_A = T_B =300K$.  From Eq. (\ref{eq:tequilT}) this is the state with the most ways to distribute the kinetic energy. Now let a sufficient amount of heat $\delta q$ move spontaneously from $A$ to $B$ such that $B$ is now one millionth of a degree hotter than $A$, so $\delta T = 10^{-6} K, T_B = 300 + 0.5\times 10^{-6}K$,  $T_A = 300 - 0.5\times 10^{-6}K$.  The amount of heat moved is given by $\delta q = m C_p \delta T/2$, where $m=1g$ and $C_p = 4.18\,\textrm{J/g/K}$ is the specific heat capacity of water, which gives $\delta q  = 2.09 \,\mu$J.  Now as the heat moves the temperatures of $A$ and $B$ change, so their average temperatures are $300 - 0.25 \times 10^{-6}K$ and $300 + 0.25 \times 10^{-6}K$ during the transfer, respectively. Using these values in Eq. (\ref{eq:tequilT}) the total number of ways to partition the kinetic energy among the atoms decreases by a factor of about  $10^{3.6 \times 10^{5}}$, which is a vast number: 1 followed by more than three hundred and fifty thousand zeros. So the probability of a spontaneous flow of heat from cold to hot producing just one millionth of a degree temperature difference is one over this vast number, practically zero. This is in spite of the small amounts of materials and a difference in temperature at the limit of measurement. This example drives home the fact that for any macroscopic situation the zeroth and second laws of thermodynamics are effectively absolute.
 
\section{Discussion}
Starting with the classical Hamiltonian description of a thermodynamic system it is shown that a concise, self-contained and self consistent derivation of the core concepts of statistical mechanics follows, with no reliance on classical (continuum) thermodynamics. The primary condition for thermal equilibrium is that the system be in the macrostate of maximum likelihood, the one that can be realized by the most microstates, which is equivalently the state of maximum entropy defined in statistical mechanics terms by Boltzmann's equation (Eq. \ref{eq:slnw2}). When the equilibrium state of maximum entropy is achieved, the mean kinetic energy per atom $\bar{E}_k$ is the same within all the parts of the system that are in thermal contact, although no \textit{a priori} equalization of any quantity was assumed in obtaining this result. Temperature is defined purely in statistical mechanics terms, from which it follows that the temperature is simply proportional to the mean kinetic energy per atom. Thus $T$ is equalized when $\bar{E}_k$ is. Moreover, it can be seen that the second law of thermodynamics is actually logically prior to the zeroth law and that the latter is not needed as a postulate. 

Within the macrostate of maximum likelihood spatial configurations follow the Boltzmann distribution, namely they are distributed exponentially with respect to their potential energy. The Boltzmann factor $exp(-U_i/k_bT)$  summed over the available spatial configurations gives the configuration partition function, Eq. (\ref{eq:partition1}), which is proportional to the total number of spatio-kinetic microstates. These relations all emerge as a straightforward consequence of classical particles undergoing random motion according to the laws of classical dynamics.

Regarding the generality of the Boltzmann distribution derivation there are several additional points. First, this distribution was derived for a thermally isolated body of atoms in a rigid container. These boundary conditions correspond to the $NVE$ or microcanonical ensemble in the standard approach. However, if this body is in contact with a much larger body with which it can exchange heat (i.e. a thermal reservoir) this does not affect the argument for the following reason. The difference in potential energy of the atoms between any two spatial configurations results in a difference in kinetic energy $\delta E_k = - \Delta U$. For constant E conditions the number of ways to distribute the kinetic energy changes according to Eq. (\ref{eq:boltz1}).  This depends only on $\delta E_k$ and the mean kinetic energy ($T$) of the atoms, not the number of atoms. The change in $W$ will be the same whether the heat evolved due a change in potential energy is shared among just the $N$ atoms in the body ($E$ is constant), or whether it is shared among them plus the atoms of the heat reservoir (corresponding to the $NVT$ or canonical ensemble conditions). Ultimately one can always consider the system plus the reservoir as a closed system at constant $E$.  \cite{callen_thermodynamics_1985} The straightforward extension to constant pressure conditions is considered in the Appendix. Second, the role of forces in the Boltzmann distribution is quite general. They act the same way to convert between potential and kinetic energy whether they are external, e.g. from gravity or an electrostatic field, or internal, from forces between atoms. Moreover, for internal forces there is no distinction between inter-molecular forces like the  van der Waals interaction and intra-molecular forces as long as the potential energy can be written in the classical Hamiltonian form of Eq. \ref{eq:Upot}. All atoms are just moving over a continuous 3N-dimensional potential energy surface according to Newtonian dynamics.

If one starts with a classical Hamiltonian then of course some systems that exhibit specific quantum properties cannot be treated: This includes systems at extremely low temperatures, or where quantum efffects are large, including liquid Helium, systems in magnetic fields, nuclear spin lattices. However, there are large areas of applied statistical mechanics and molecular simulations where, due to the system’s complexity, one has to start from a classical Hamiltonian. This includes many systems one wishes to study in chemical physics and biophysics such as polymers, colloids, membranes and macromolecules. Here there is no possibility of starting from a quantum treatment and taking the classical limit. So there is no downside and some advantages to starting with a classical a treatment. For one, the effect of system size can easily be examined. Here it is shown that there is a surprisingly rapid approach to the macroscopic limit for entropy changes, which is reached with just $10^3-10^4$ atoms.

This approach certainly cannot account for the negative temperatures claimed for specially prepared non-equilibrium systems \cite{braun_negative_2013}. In the classical treatment, temperature is proportional to  the mean kinetic energy per atom, which is always is always a positive quantity. But in the author’s judgement the impossibility of $T<0$  is not a limitation of the classical treatment, but a failure of the concept of negative temperature. In cases where negative temperatures - as defined thermodynamically -  are claimed,  these are just special non-equilibrium cases where states of higher energy are more populated than lower energy states. Using the term negative temperature actually has less explanatory power (but more shock value!) than simply saying that the populations of energy levels are temporarily inverted and it is inconsistent with statistical mechanics definitions of heat capacity. 

\begin{acknowledgments}
I would like to acknowledge my long time collaborator and late colleague, Dr. Franz Matschinsky. Extended discussions with Franz over the years have greatly enriched my work and stimulated the writing of this paper. 
\end{acknowledgments}

\appendix*   
\section{} 

\subsection{Effect of simplifying assumptions}
\label{sub:masseffect}
For clarity at two points simplying assumptions were made. First,  the scaling law Eq. (\ref{eq:ways_Natom}) was derived for atoms of identical masses. If atoms have different masses the surface of constant energy traced out by the $3N$\nbd-dimensional velocity vectors is a $3N$\nbd-dimensional hyper-ellipsoid whose ratios of semi-axis lengths along the $i$\nbd-atom and $j$\nbd-atom velocity directions are given by $(m_i/m_j)^{1/2}$. Since the shape of this hyper\nbd-ellipsoid is fixed by the mass ratios its surface area still scales as $E_k^{(3N-1)/2}$, merely with a different (and very complicated) numerical factor. Since we only need ratios of numbers of ways to distribute kinetic energy this factor cancels, again giving Eq. (\ref{eq:ways_Natom}). Regarding mass effects, when calculating the absolute or $3^{rd}$ law entropy of an ideal gas using quantum mechanics (the Sackur-Tetrode approach) the final result for the \textit{absolute} entropy contains an additive mass dependent constant. \cite{mcquarrie_statistical_1976} Since this constant is a fixed property of each atom type and does not depend on its position, velocity or kinetic energy, it plays no part in the origin of the Boltzmann  distribution or changes in entropy.  Indeed, mass does not appear in the Boltzmann distribution. \cite{ hill_introduction_1962, mcquarrie_statistical_1976}

The second assumption is that for mathematical convenience a reference state set of atomic positions with $U=0$ is used in Eq. (\ref{eq:waysref}).  A reference state with non\nbd-zero potential energy $U_0$ could be used instead. In this case the constant factor in front of the partition function sum is just $W_0 \exp(-U_0/k_b T)$, but the sum itself is unchanged. 
\subsection{Generality of Eq. (\ref{eq:waystotal1})}
This equation gives the change in the number of ways to distribute kinetic energy for a small addition of heat, $\delta q << E_k$ to a body, regardless of the composition of the body or whether there are forces acting between the atoms. Note, however, the difference between an ideal gas and the general case as more and more heat is added. For an ideal gas there is no potential energy, only kinetic energy. So the latter increases by exactly the amount of heat added. For the more general case as the temperature rises the spatial configurations with higher potential energies will be increasingly populated, as shown by the Boltzmann factor, Eq. (\ref{eq:boltz3}). As a consequence the average potential energy will rise (\textit{cf.} Eq. (\ref{eq:averages})). This means some of the added heat is converted to potential energy and less remains as kinetic energy to raise the temperature. In other words when there are interactions between atoms the heat capacity will be greater than that of an ideal gas. In both cases, though, Eq. (\ref{eq:waystotal1}) gives the \textit{rate} of change of $W_{tot}$ with respect to $\delta q$ at any given temperature, which is all that is needed to establish the conditions for thermal equilibrium.
\subsection{Accounting for volume changes}
\label{sub:dvolume}
Constant pressure conditions are more common experimentally than constant volume conditions.  It is straightforward to account for the effects of volume changes. Suppose upon moving between spatial configurations $A$ and $B$ (Fig. \ref{fig:sphereforce}) the volume of the system changes by $\Delta V = V_B - V_A$. Now some of the change in potential energy is converted to work against the external pressure $P \Delta V$.  Only the remainder is converted to heat. Thus $\delta q = - \Delta U - P \Delta V$.  Terms  $\Delta U+P \Delta V$ and $U_i + P V_i$ now replace $\Delta U$ and $U_i$, respectively, in Eqs. (\ref{eq:wayratio2}) - (\ref{eq:waystotal}). The macroscopic thermodynamic average quantity obtained from Eq. (\ref{eq:averages}) with $Y_i = U_i + P V_i$ is now $H=U+PV$, the enthalpy.
\newpage
\bibliography{ZerothLaw}
\bibliographystyle{pnas2009.bst}
\end{document}